\documentclass[article,twocolumn,superscriptaddress,frontmatterverbose,amsmath,amssymb,prx,numbers]{revtex4-2}

\usepackage[english]{babel}
\usepackage{amssymb,amsthm,amsmath,setspace}
\usepackage{graphicx}
\usepackage[dvips]{epsfig}
\usepackage{xcolor,colortbl, multirow}
\usepackage{hyperref}
\addto\extrasenglish{%
}
\usepackage{xcolor, soul} 

\usepackage[version=3]{mhchem} 
\usepackage{booktabs}
\usepackage{multirow}
\usepackage{siunitx}
\usepackage{longtable}
\usepackage{comment}

\begin{document}

\title{Does the total energy difference method for modelling core level photoemission fail for bigger molecules?}
\author{Marta Berholts}
\affiliation{Department of Physics, University of Tartu, EST-50411 Tartu, Estonia}
\author{Tanel Käämbre}
\affiliation{Department of Physics, University of Tartu, EST-50411 Tartu, Estonia}
\author{Arvo Tõnisoo}
\affiliation{Department of Physics, University of Tartu, EST-50411 Tartu, Estonia}
\author{Rainer Pärna}
\affiliation{Department of Physics, University of Tartu, EST-50411 Tartu, Estonia}
\author{Vambola Kisand}
\affiliation{Department of Physics, University of Tartu, EST-50411 Tartu, Estonia}
\author{Juhan Matthias Kahk}
\affiliation{Department of Physics, University of Tartu, EST-50411 Tartu, Estonia}
\date{\today}
\begin{abstract}
The $\Delta$-Self-Consistent-Field ($\Delta$SCF) method permits calculations of core electron binding energies in materials and molecules at a modest computational cost.  However, it has been reported that whilst this method works well for small molecules, its accuracy drops off dramatically when larger systems are considered. Particularly large errors have been reported for the anthrone molecule, which consists of 25 atoms. In this work, the gas-phase photoelectron spectrum of anthrone is revisited both computationally and experimentally. The measured C 1s binding energies in anthrone differ markedly from previously published values, and the new experimental results are in good agreement with $\Delta$SCF calculations based on the SCAN functional. In addition, the performance of the $\Delta$SCF method is evaluated for a dataset of 44 core electron binding energies from medium sized molecules containing between 10 and 40 atoms. The mean absolute error for this dataset – 0.19 eV – is comparable to the results of previous computational benchmarks. Overall, these results and general theoretical considerations indicate that the $\Delta$SCF method is suitable for modelling localized excitations in both small and large molecules, and applications to other extended systems are also promising.
\end{abstract}
 
\maketitle
\section{Introduction}

The total energy difference method based on Density Functional Theory (DFT) has recently attracted significant attention as a computationally affordable approach to modelling electronic excitations that would be challenging to tackle otherwise \cite{hait_excited_2020,hait_orbital_2021,kahk_combining_2023,hall_characterizing_2023,nishino_identification_2024,carter-fenk_state-targeted_2020,paetow_excited_2025,pollack_scf_2025}. In the context of X-ray spectroscopies, it is commonly known as the $\Delta$-Self-Consistent-Field ($\Delta$SCF) method \cite{bagus_self-consistent-field_1965,banna_accurate_1982,triguero_separate_1999,besley_self-consistent-field_2009,bagus_extracting_2018,kahk_accurate_2019,hait_highly_2020,kahk_core_2021,klein_nuts_2021,besley_modeling_2021,yang_foundation_2024}. In a $\Delta$SCF calculation, the excitation energy is obtained as the total energy difference between the ground state and the final state with a core hole. For the final state, a DFT calculation with a non-Aufbau-principle occupation of the Kohn-Sham eigenstates is performed. The $\Delta$SCF approach has been shown to yield highly accurate energies of charged and neutral excitations in small molecules, when compared against experimental data from X-ray Photoelectron Spectroscopy (XPS) and X-ray Absorption Spectroscopy (XAS) \cite{pueyo_bellafont_performance_2016,kahk_accurate_2019,hait_highly_2020}. It has also been demonstrated how this appoach can be extended to various kinds of systems, including molecules with unpaired electrons \cite{hait_accurate_2020} and periodic solids \cite{kahk_core_2021,kahk_combining_2023}.

In a more general sense, the strategy of targetted convergence of the Kohn-Sham equations towards a specific excited electronic configuration has been termed Orbital-Optimized excited state DFT (OO-DFT) \cite{hait_orbital_2021}. Besides core excitations, OO-DFT has also been shown to accurately describe charge transfer excitations in molecules \cite{hait_orbital_2021}, as well as certain types of doubly excited states \cite{hait_excited_2020}.

The justification for performing DFT calculations with non-Aufbau-principle occupations has been, until recently, mostly empirical. The $\Delta$SCF method has been pursued because it gives good results, even though it is understood that the Hohenberg-Kohn theorem does not hold for excited states, i.e. there is no one-to-one correspondence between an excited state charge density and the external potential. However, Yang and Ayers have recently provided a formal justification for the $\Delta$SCF method in reference \cite{yang_foundation_2024}, and shown that if, for instance, the noninteracting wavefunction, $\Psi$, is used as the fundamental descriptor of a system, then the same universal functional of $\Psi$ will give the exchange-correlation energy for both the ground state and excited state. This also justifies the current practice of using approximate ground state exchange correlation functionals in $\Delta$SCF calculations.

Still, other questions about the general usefulness of the $\Delta$SCF approach remain. A particularly pertinent issue is the one of size-extensivity. With standard exchange-correlation functionals, when calculating the first ionization energy, as the system size increases, the $\Delta$SCF result simply tends towards the orbital eigenvalue of the highest occupied state. This has been demonstrated for fragments of polyacetylene in ref. \cite{pinheiro_length_2015}, and a similar trend is known to hold for periodic systems \cite{corsetti_system-size_2011,persson_n_2005}. In other words, it becomes meaningless to use the DFT – $\Delta$SCF method for calculating the 1st ionization energy of an extended system, as it provides no additional information compared to a simple ground state DFT calculation. 

However, it has already been shown that the same does not hold for $\Delta$SCF calculations of core electron binding energies, in which the core hole must always be localized at a single atom, regardless of system size \cite{kahk_combining_2023}. That is to say, 

\begin{equation}
\lim\limits_{n \to \infty} (E_{N-1,\,\mathrm{core\:hole}}(n) - E_{N,\,\mathrm{ground}}(n)) \neq -\epsilon_\mathrm{core},
\label{eqn1}
\end{equation}

where $N$ is the number of electrons and $n$ is the number of atoms, $E_{N,\,\mathrm{ground}}(n))$ is the ground state total energy, $E_{N-1,\,\mathrm{core\:hole}}(n)$ is the total energy of the $N-1$ electron final state with a core hole, and $\epsilon_\mathrm{core}$ is the Kohn-Sham eigenvalue of the core orbital.

The key aspect here is that the creation of a localized core hole always brings about a strong screening response, i.e. the remaining electrons will rearrange in response to the change in the local potential in the vicinity of the atom with a missing core electron, regardless of system size. As such, in calculations of core electron binding energies, there is no inherent reason for why the $\Delta$SCF method should be destined to fail when applied to large molecules or other extended systems.

\begin{figure*}[ht]
 \centering
  \includegraphics[width=400pt]{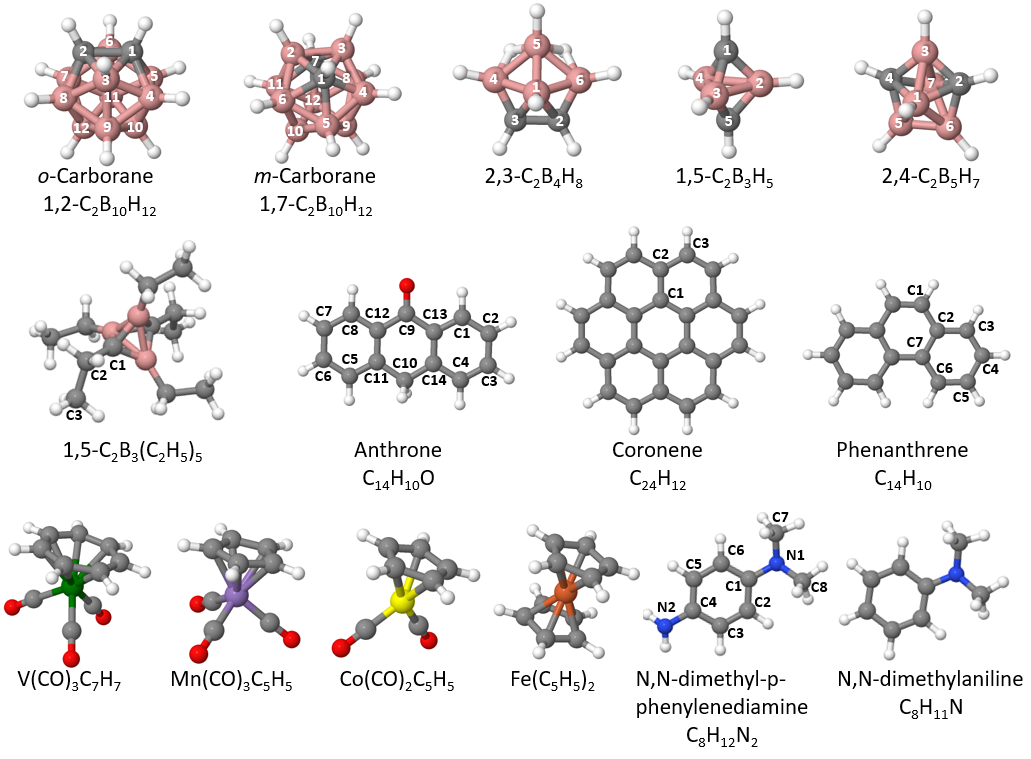}
  \caption{Molecular set used in this work. Shown structures are geometry-optimized. Color code: grey, C; white, H; red, O; blue, N; green, V; orange, Fe; violet, Mn; pink, B; yellow, Co. The molecules are labeled to support notations in \autoref{tab:comparison}.}
  \label{fig:Structures}
 \end{figure*}

Nevertheless, empirically, it has been observed that the same $\Delta$SCF formalism that produced good results for small molecules gave much larger errors when applied to somewhat larger organic molecules containing fused aromatic rings. In particular, in ref. \cite{golze_core-level_2018} it was shown that whilst $\Delta$SCF calculations based on the PBE0 hybrid functional reproduced the C 1s binding energies in methane, dimethyl ether, and hydrogen cyanide with acceptable accuracy (errors of approximately 0.5 eV), applying the same method to anthrone and phenanthrene (structures shown in Figure \ref{fig:Structures}) resulted in significantly larger errors of 0.7–1.5 eV. These findings have been interpreted as indicative of a general tendency of the $\Delta$SCF method to perform poorly when applied to more complex molecules \cite{golze_accurate_2020,golze_accurate_2022,li_benchmark_2022,mukatayev_electron_2022}.

In order to investigate this issue further, we re-examined the core level photoelectron spectrum of anthrone using both experimental and computational tools. We have also performed additional theoretical calculations for a dataset of 44 experimental core electron binding energies in 15 molecules containing between 10 and 40 atoms. The dataset contains molecules from several different classes, including organometallic compounds, carborane clusters, and polyaromatic systems – this selection has been largely dictated by the availability of reference data from gas phase photoelectron spectroscopy. The rest of this manuscript is organized as follows: the experimental and computational methods are described next, then, the photoemission spectrum of anthrone is analyzed in detail, and finally, the accuracy of the DFT-$\Delta$SCF method based on the SCAN functional, employing the same numerical settings as in references \cite{kahk_accurate_2019,kahk_core_2021,kahk_predicting_2022,kahk_combining_2023}, is evaluated for the new dataset of medium-sized molecules.

\section{Methods}
\label{sec:methods}
All reported DFT calculations were performed using the all-electron electronic structure program package FHI-aims \cite{blum_ab_2009,abbott_roadmap_2025,yu_elsi_2018,havu_efficient_2009}. The strongly constrained and appropriately normed (SCAN) exchange correlation functional \cite{sun_strongly_2015}, implemented via dfauto \cite{strange_automatic_2001} was employed throughout. The structures of the molecules were relaxed using the trust-radius enhanced BFGS algorithm \cite{nocedal_numerical_2006,broyden_convergence_1970,fletcher_new_1970,goldfarb_family_1970,shanno_conditioning_1970} until the forces on all atoms were below 5 meV/\AA. For geometry optimization, the FHI-aims default “tight” basis sets and numerical integration grids were used on all atoms \cite{blum_ab_2009,abbott_roadmap_2025}. Core electron binding energies were calculated using the $\Delta$SCF method as the total energy difference between the N-electron ground state and the (N-1)-electron final state with a core hole. In the final state calculations, a core hole was introduced by fixing the occupancy of a specific Kohn-Sham state in the spin-up spin channel to zero. For the atom with a core hole, a larger basis set was used, to facilitate the relaxation of the remaining core and valence electrons in the presence of the core hole. The core-hole-adapted basis sets used in this work are given in the supplementary information.

The GW calculation of the valence level spectrum of anthrone was also performed in FHI-aims. Quasiparticle energy levels of the valence states were obtained using a “one-shot” G$_0$W$_0$ calculation based on a PBE0 \cite{adamo_toward_1999} starting point. In the implementation used in this work \cite{ren_resolution--identity_2012}, the self-energy is first calculated on the imaginary frequency axis, and then analytically continued onto the real axis. For the analytic continuation, a Pade approximant with 16 parameters was employed. Quadruple-zeta numerical atom-centered basis sets with valence-correlation consistency from reference \cite{zhang_numeric_2013} were used. The simulated valence band photoelectron spectrum was constructed according to the Gelius approximation \cite{gelius_esca_1972}, i.e., as a cross-section weighted density of states (DOS) curve. Per-electron photoionization cross-sections corresponding to hv = 50 eV were obtained from ref. \cite{yeh_atomic_1985} – the numerical values of the used cross-sections are as follows: C 2s: 0.49, C 2p: 0.58, O 2s: 0.44, O 2p: 1.29.

The photoelectron spectroscopy (PES) measurements were performed at the gas phase end-station of the FinEstBeAMS beamline \cite{parna_finestbeams_2017,kooser_gas-phase_2020,chernenko_performance_2021} of the MAX IV synchrotron facility in Lund, Sweden \cite{tavares_max_2014,martensson_saga_2018,robert_max_2023}. Prior to the photoemission measurements, a Wiley-McLaren type time-of-flight ion spectrometer equipped with a Roentdek microchannel plate and a HEX-anode detector was used to record a mass spectrum of the vapour present in the analysis chamber in order to ensure that the molecular ion is detected and that the vacuum chamber is free from possible contamination. The recorded mass spectra are included in the supplemental information. Anthrone was purchased from Sigma-Aldrich with a stated purity of 97\%. For measurement, anthrone vapours were introduced into the analysis chamber by heating the solid powder using an MBE-Komponenten effusion cell. To achieve the required sample density at the interaction region the solid was heated to 50$^\circ$C. Core and valence level photoelectron spectra were recorded using the Scienta R4000 hemispherical electron analyzer equipped with a fast microchannel plate and a resistive anode (Quantar Inc.) detector. For the C 1s measurement at h$\nu$ = 350 eV and for the O 1s measurement at h$\nu$ = 600 eV, the pass energy of the analyzer was set to 200 eV and the analyzer slit was set to 0.9 mm. For the valence photoelectron spectrum recorded at h$\nu$ = 50 eV, an analyzer pass energy of 50 eV and an analyzer slit of 0.3 mm were used. The calibration of the XPS spectrum was performed using the Ar 2p$_{3/2}$ photoelectron line at 248.63 eV \cite{avaldi_near-threshold_1994,jolly_core-electron_1984}. The binding energy scale of the valence level spectrum was calibrated using the 1b$_1$ peak of H$_2$O vapour at 12.62 eV \cite{karlsson_isotopic_1975,bock_k_1982}. The energy resolution, as obtained from fitting Gaussian curves to Ar 2p$_{3/2}$ and Ar 3p$_{3/2}$ photoelectron lines, was 540 meV, 900 meV, and 109 meV at full width half maximum (FWHM) for the C 1s, O 1s, and valence measurements, respectively. 

\section{Results}

\begin{figure*}[ht]
 \centering
  \includegraphics[width=430pt]{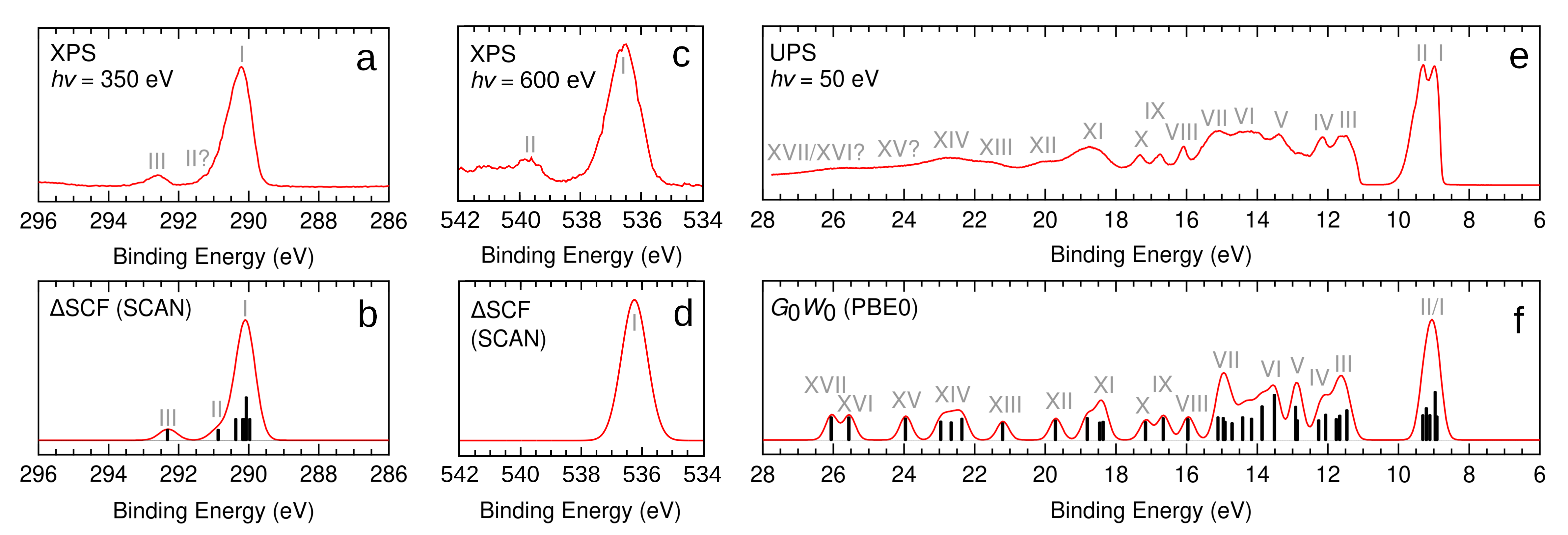}
  \caption{Experimental and calculated XPS spectra of anthrone showing the C 1s region: a) and b), and the O 1s region: c) and d). Experimental and calculated valence photoelectron spectra of anthrone are shown in panels e) and f). Traces of the spectrum of water were subtracted from the experimental valence spectrum.}
  \label{fig:Anthrone_XPS_Valence}
 \end{figure*}

The C 1s, O 1s, and valence regions of the photoelectron spectrum of anthrone are shown in Figure \ref{fig:Anthrone_XPS_Valence}. The experimental C 1s spectrum (Figure \ref{fig:Anthrone_XPS_Valence}a) contains two main features: a strong, broad, and asymmetric peak at 290.27 eV, and a weaker peak at 292.7 eV. The measured peak positions are significantly different from the values reported in ref. \cite{jolly_pi-donor_1976,bakke_table_1980}: 290.72 eV and 293.6 eV, respectively (see the comparison in Table \ref{tab:anthrone_BE}). Due to this discrepancy, the method used for the calibration of the absolute binding energy scale in this study is described in detail next.

With anthrone vapour already present in the analysis chamber, Ar gas was allowed to enter the chamber via a leak valve. The Ar 2p$_{3/2}$ binding energy was measured immediately before the C 1s spectrum using the same beamline and analyzer settings, without adjusting either of the two (except for the detected kinetic energy range) between the two measurements. When measuring the C 1s region, no drift in energy between consecutive analyzer sweeps was detected. In addition, a separate measurement of the C 1s region was also performed at h$\nu$ = 600 eV. From that measurement (with a similar procedure for the calibration of the energy scale) peak positions of 290.37 eV and 292.76 eV for the strong and weak peaks are obtained. We therefore conclude that the energy positions of the strong and weak peaks have been determined as 290.3 eV and 292.7 eV with an accuracy of approximately $\pm$0.1 eV. 

 \begin{table*}[t]
\small
\caption{Experimental core electron binding energies (BE) of anthrone at different ionization sites compared to literature values. The BEs are given in eV. The difference is calculated as BE(own)-BE(ref.).}\label{tab:anthrone_BE}
\centering \vspace{0.05in}
\begin{tabular}{lccc} \toprule
                  \textbf{core level}     & \textbf{BE (own)} & \textbf{BE (ref.)} & \textbf{Difference}\\ \midrule
                 C 1s (C1-C8, C11-C14)     & 290.3      & 290.72\cite{bakke_reversal_1977,jolly_core-electron_1984} & -0.42\\
                 C 1s (C10)                & 290.9      & -   & -\\
                 C 1s (C9)                 & 292.7      & 293.6\cite{bakke_reversal_1977,jolly_core-electron_1984}  & -0.9\\
                 O 1s                      & 536.6      & 536.89\cite{bakke_reversal_1977,jolly_core-electron_1984} & -0.29\\ \midrule
\bottomrule                          
\end{tabular}             
\end{table*}

The experimental O 1s spectrum of anthrone (Figure \ref{fig:Anthrone_XPS_Valence}c) contains one main peak centered at 536.6 eV. In addition, a weak peak at $\sim$ 539.8 eV is attributed to residual H$_2$O vapour present in the analysis chamber, and possibly to satellite features of the main anthrone peak. The experimental valence level spectrum (Figure \ref{fig:Anthrone_XPS_Valence}e) contains a number of features, with the first peak corresponding to ionization from the highest energy molecular orbital (HOMO) at 9.0 eV. Again, due to the presence of a small amount of water vapour in the chamber, the measured valence spectrum contained minor contributions from water – these have been subtracted from the spectrum shown in Figure \ref{fig:Anthrone_XPS_Valence}e – details are provided in the supplementary information.

Theoretical C 1s, O 1s, and valence level spectra of anthrone are shown in panels b, d, and f of Figure \ref{fig:Anthrone_XPS_Valence}. The calculated C 1s and O 1s spectra have been constructed as a sum of gaussian functions of fixed height and width (FWHM of 0.6 eV for C 1s and 1.0 eV for O 1s) centered at the binding energies from the $\Delta$SCF calculations. The calculated valence band spectrum has been obtained by applying constant gaussian broadening with FWHM = 0.4 eV to the cross-section weighted density of states (see Section \ref{sec:methods}). It is emphasized that no empirical shifts have been applied – Figure \ref{fig:Anthrone_XPS_Valence} represents a direct comparison of absolute calculated core and valence electron binding energies with absolute experimental binding energies referenced to the vacuum level.

Figure \ref{fig:Anthrone_XPS_Valence} shows that the theoretical spectra are in overall excellent agreement with the experimental data. In particular, the predicted lineshape and position of the theoretical C 1s spectrum match experiment very closely, indicating that the $\Delta$SCF method works well in this instance. Based on the calculations, the C 1s spectrum of anthrone consists of three distinct features. Feature I at 290.11 eV (290.3 eV in the experiment) corresponds to all sp$^2$ carbons with C-C and C-H bonds. Feature II is a weak shoulder on the high binding energy side of feature I, originating from the sp$^3$ carbon (C10 in Figure \ref{fig:Structures}). The calculated C 1s binding energy of C10 is 290.85 eV and although this feature is not well resolved in the experimental spectrum, the asymmetric shape the main peak in Figure \ref{fig:Anthrone_XPS_Valence}a is consistent with the presence of Feature II at approximately 290.9 eV. Feature III corresponds to the carbon atom next to the oxygen, and appears at 292.28 eV in the theoretical spectrum and at 292.7 eV in the experiment. The calculated O 1s binding energy (536.25 eV) is also close to the experimental value of 536.6 eV.

The calculated valence level spectrum of anthrone is also very similar in appearance to the experimental result. The most notable difference between theory and experiment is the appearance of peaks I and II that are well resolved in the experiment, but merged together in the calculation. A closer examination of the theoretical results indicates that a splitting of the valence orbital energies between 9 and 10 eV into two groups is present, but the magnitude of this splitting is underestimated, meaning that the peaks merge together when broadening is applied. It is also observed that, as expected, in the experimental spectrum the peaks corresponding to the deeper valence levels appear smeared out due to lifetime broadening.

\begin{table*}
\small
\caption{Comparison of experimental and theoretical core electron binding energies (BE) of medium-sized gas-phase molecules at different ionization sites. BEs were calculated using the $\Delta$SCF method. Given error is the difference between the computationally and experimentally determined BE.}\label{tab:comparison}
\centering \vspace{0.05in}
\begin{tabular}{llccc} \toprule
                            & \textbf{core}  & \textbf{theor. BE}      & \textbf{exp. BE}       & \textbf{error} \\ 
     \textbf{molecule}             & \textbf{level} & \textbf{(eV)} & \textbf{(eV)} & \textbf{(eV)} \\\midrule
  1,5-\ce{C2B3(C2H5)5}      & B 1s                   & 194.51    & 194.7 \cite{allison_molecular_1972, jolly_core-electron_1984}                            & -0.19  \\
                            & C 1s (C1)              & 288.8     & 287.2                                                                                    & 1.60  \\
                            & C 1s (C2)              & 289.64    & 289                                                                                      & 0.64  \\
                            & C 1s (C3)              & 290.15    & 290.3                                                                                    & -0.15  \\ \midrule
  1,5-\ce{C2B3H5}           & B 1s (B2-B4)                   & 195.94    & 196 \cite{allison_molecular_1972, jolly_core-electron_1984}                              & -0.06  \\
                            & C 1s (C1, C5)                   & 290.16    & 290.2                                                                                    & -0.04  \\\midrule
  2,3-\ce{C2B4H8}           & B 1s (B4, B6)               & 195.17    & 195.2 \cite{allison_molecular_1972}                                                      & -0.03  \\
                            & B 1s (B5)                   & 195.92    & 196                                                                                    & -0.08  \\ 
                            & B 1s (B1)                   & 195.47    & 195.7     (assignments swapped)                                                                                 & -0.23  \\
                            & C 1s (C2, C3)                  & 290.83    & 290.9                                                                                    & -0.07  \\\midrule
  2,4-\ce{C2B5H7}           & B 1s (B5, B6)               & 194.84    & 194.9 \cite{allison_molecular_1972, jolly_core-electron_1984}                            & -0.06  \\ 
                            & B 1s (B3)                   & 195.6     & 195.6                                                                                    & 0.00 \\
                            & B 1s (B1, B7)               & 196.15    & 196.1                                                                                    & 0.05  \\
                            & C 1s (C2, C4)                  & 291.09    & 291                                                                                      & 0.09 \\ \midrule
 1,7-\ce{C2B10H12}          & B 1s (B9, B10)              & 194.78    & 194.7 \cite{allison_molecular_1972, jolly_core-electron_1984}                            & 0.08 \\
                            & B 1s (B5, B12)              & 195.33    & 195.4                                                                                    & -0.07  \\ 
                            & B 1s (B4, B6, B8, B11)      & 195.47    & 195.6                                                                                    & -0.13  \\
                            & B 1s (B2, B3)               & 196.13    & 196.4                                                                                    & -0.27  \\ \midrule
 1,2-\ce{C2B10H12}          & B 1s (B9, B12)              & 194.66    & 194.5 \cite{allison_molecular_1972, jolly_core-electron_1984}                            & 0.16  \\
                            & B 1s (B8, B10)              & 194.81    & 195                                                                                      & -0.19 \\ 
                            & B 1s (B4, B5, B7, B11)      & 195.5     & 195.6                                                                                    & -0.10  \\
                            & B 1s (B3, B6)               & 196.17    & 196.3                                                                                    & -0.13  \\ \midrule
N,N-dimethyl-p-             & C 1s (C1, C4, C7, C8)             & 290.78    & 290.93 \cite{jolly_core-electron_1984}                                           & -0.15  \\ 
phenylenediamine            & C 1s (C2, C3, C5, C6)             & 289.46    & 289.66                                                                                   & -0.20  \\
                            & N 1s (N1, N2)                  & 404.84    & 404.84                                                                                   & 0.00  \\ \midrule
 N,N-dimethylaniline        & N 1s                   & 404.97    & 405.085 \cite{cavell_site_1977, brown_determination_1980, jolly_core-electron_1984}      & -0.115 \\ \midrule
 phenanthrene               & C 1s (C2,C7)           & 289.95    & 290.2 \cite{fronzoni_vibrationally_2014}                                                 & -0.25 \\
                            & C 1s (C1, C3-6)        & 289.65    & 289.9                                                                                    & -0.25  \\ \midrule
 coronene                   & C 1s (C1,C2)           & 289.7     & 289.9 \cite{fronzoni_vibrationally_2014}                                                 & -0.20  \\
                            & C 1s (C3)              & 289.41    & 289.6                                                                                    & -0.19  \\ \midrule
 anthrone                   & C 1s (C1-C8, C11-C14)  & 290.11    & 290.3 (own data)                                                                         & -0.19  \\
                            & C 1s (C10)             & 290.85    & 290.9                                                                                    & -0.05 \\
                            & C 1s (C9)              & 292.28    & 292.7                                                                                    & -0.42 \\
                            & O 1s                   & 536.25    & 536.6                                                                                    & -0.35  \\ \midrule
 \ce{Co(CO)2C5H5}           & C 1s (\ce{C5H5})            & 290.78    & 290.67 \cite{xiang_x-ray_1982, jolly_core-electron_1984}                                 & 0.11  \\
                            & C 1s (\ce{CO})              & 292.68    & 292.68 \cite{avanzino_study_1980, jolly_core-electron_1984}                              & 0.00  \\
                            & O 1s                   & 538.92    & 539.01 \cite{avanzino_study_1980, jolly_core-electron_1984}                              & -0.09  \\ \midrule
 \ce{Mn(CO)3C5H5}           & C 1s (\ce{C5H5})            & 290.94    & 290.93 \cite{chen_x-ray_1981, calabro_effects_1981, jolly_core-electron_1984,avanzino_study_1980} & 0.01 \\ 
                            & C 1s (\ce{CO})              & 292.6     & 292.53                                                                                   & 0.07  \\
                            & O 1s                   & 538.82    & 538.875                                                                                  & -0.05  \\ \midrule
 \ce{Fe(C5H5)2}             & C 1s                   & 289.9     & 290.03 \cite{bakke_reversal_1977, jolly_core-electron_1984}                              & -0.13  \\ \midrule 
 \ce{V(CO)3C7H7}            & C 1s (\ce{C7H7})            & 290.62    & 290.4 \cite{rietz_x-ray_1975, jolly_core-electron_1984}                                  & 0.22  \\
                            & C 1s (\ce{CO})              & 291.71    & 291.9                                                                                    & -0.19  \\
                            & O 1s                   & 538.28    & 538.8                                                                                    & -0.52 \\ \midrule 
\multicolumn{5}{l}{Mean absolute error = 0.19 eV (0.15 eV) with (without) the outlier included.}\\ 
\bottomrule                          
\end{tabular}             
\end{table*} 

A numerical comparison of experimental and calculated core electron binding energies in anthrone and 14 other medium sized molecules is presented in Table \ref{tab:comparison}. Overall, we observe excellent agreement between theory and experiment, with a mean error of -0.05 eV and a mean absolute error of 0.19 eV for the whole dataset. It is noted that for 2,3-C$_2$B$_4$H$_8$ the assignments of two boron atoms, B1 and B5 (see Figure \ref{fig:Structures} for the numbering scheme) has been swapped, compared to the original assignments in reference \cite{allison_molecular_1972}. Both environments correspond to exactly one B atom, meaning that the assignment cannot be determined from the experimental core level spectrum alone. In reference \cite{allison_molecular_1972}, the peaks were assigned based on the semiempirical calculations available at the time, however, if the theoretical values from this work are taken as a guide, swapping the assignments significantly reduces the errors between theory and experiment from +0.22 eV and --0.53 eV to --0.08 eV and --0.23 eV for B5 and B1, respectively. We also note that a significant part of the average error for the whole dataset originates from just one datapoint – C1 in 1,5-C$_2$B$_3$(C$_2$H$_5$)$_5$ – where the calculated result differs from the experimental value by 1.6 eV. The origin of this very large error is presently not understood, as otherwise the agreement between theory and experiment in carborane clusters appears satisfactory.  C1 in 1,5-C$_2$B$_3$(C$_2$H$_5$)$_5$ is an unusual case as the reported binding energy of 287.2 eV is very low for a gas phase C 1s peak. If this datapoint is excluded, the mean absolute error for the dataset reduces to 0.15 eV, and the mean error becomes -0.09 eV.

\section{Summary}

In this work the suitability of the $\Delta$SCF method for predicting core electron binding energies in medium-sized molecules containing between 10 and 40 atoms has been examined. Overall, we have found the method to perform very well – the mean absolute error (0.19 eV) and mean error (-0.05) from this study are comparable to the values obtained in previous works where the same theoretical approach was used \cite{kahk_accurate_2019,kahk_core_2021,kahk_predicting_2022,kahk_combining_2023}. We have not found any evidence to suggest that the accuracy of calculated binding energies degrades with increasing molecular size. The large errors reported for the anthrone molecule in a previous theoretical work are attributed to a problem with experimental reference values. After remeasuring the gas-phase photoelectron spectrum of anthrone, we find good agreement between the new experimental data and $\Delta$SCF calculations based on the SCAN functional.

\section{Associated content}

Supporting Information is available: photoexcited mass spectra of anthrone, the subtraction of the contribution of water vapour from the valence band photoelectron spectrum, isosurface plots demonstrating the localization of the core hole in anthrone, core-enhanced basis sets used in this work. A related dataset is available in the Zenodo repository: input and output files for the $\Delta$SCF and G$_0$W$_0$ calculations \cite{berholts_supplementary_2026}. 

\section{Acknowledgements}
This project has received funding from the European Union's Horizon Europe research and innovation programme under grant agreements no. 101131173 (BETTERXPS) and 101159716 (EXANST). JMK and MB acknowledge support from the Estonian Research Council grant number PSG1037. All authors acknowledge support from the Estonian Ministry of Education and Research grant number TK210. We acknowledge the University of Tartu, Estonia for awarding this project access to the LUMI supercomputer, owned by the EuroHPC Joint Undertaking, hosted by CSC (Finland) and the LUMI consortium through ETAIS, Estonia.

\section{References}

\end{document}